\documentclass[a4paper,11pt]{article}
\usepackage{pos}
\usepackage{bbm,color}
\usepackage{mathtools}
\usepackage{hyperref}
\usepackage{graphicx}
\usepackage{epsfig}
\usepackage{euscript}
\usepackage{xcolor}
\usepackage{fancybox}
\usepackage{enumerate}
\usepackage{widetext}
\usepackage{subfigure}
\definecolor{brightturquoise}{rgb}{0.03, 0.91, 0.87}
\definecolor{awesome}{rgb}{1.0, 0.13, 0.32}
\definecolor{armygreen}{rgb}{0.29, 0.33, 0.13}
\definecolor{aqua}{rgb}{0.0, 1.0,1.0}
\definecolor{maroon(html/css)}{rgb}{0.5, 0.0,0.0}
\definecolor{pinegreen}{rgb}{0.0, 0.47,0.44}
\definecolor{red-brown}{rgb}{0.65, 0.16,0.16}

\title{The Theory of Resummed Quantum Gravity:
Phenomenological Implications}

\author*[a]{B.F.L. Ward}

\affiliation[a]{Baylor University,\\
   Waco, TX, USA}

\emailAdd{bfl\_ward@baylor.edu}

\abstract{We present an overview of the phenomenological implications of the theory of resummed quantum gravity. We discuss its prediction for the cosmological constant in the context of the Planck scale cosmology of Bonanno and Reuter, its relationship to Weinberg's asymptotic safety idea, and its relationship to Weinberg's soft graviton resummation theorem. We also discuss constraints and consistency checks of the theory.}
\centerline{BU-HEPP-20-06, Dec., 2020}
\FullConference{%
  40th International Conference on High Energy physics - ICHEP2020\\
  July 28 - August 6, 2020\\
  Prague, Czech Republic (virtual meeting)
}

\begin{document}
\maketitle

\baselineskip=10pt
\section{Introduction}
\label{intro}
We use the well-known elementary example of ``summation'', 
\begin{equation}
\frac{1}{1-x} = \sum_{n=0}^{\infty}x^n,
\label{eq-elmtry}
\end{equation} 
to illustrate why resummation can be worth its pursuit. Even though the mathematical tests for convergence of the 
series would only guarantee convergence for $|x|$ <1, this geometric series is summed to infinity to yield the analytic result that is well-defined except for a pole at x=1. The result of the summation yields a function that is well-defined in the entire complex plane except for the simple pole at x=1 -- 
infinite order summation has yielded behavior very much improved from what one sees order-by-order in the respective series.\par
We are thus motivated to ``resum' series that are already being summed to seek improvement in our knowledge of the represented function. This we illustrate as follows:
\begin{equation}
\sum_{n=0}^{\infty}C_n \alpha_s^n \begin{cases}&= F_{\rm RES}(\alpha_s)\sum_{n=0}^{\infty} B_n \alpha_s^n,\; \text{\rm EXACT}\\
                                                                                                                                      &\cong  G_{\rm RES}(\alpha_s)\sum_{n=0}^{\infty} \tilde{B}_n \alpha_s^n,\; \text{\rm APPROX.}\end{cases}
                                                                                                                                      \label{eq-res}
\end{equation}
On the LHS (left-hand side) we have the original Feynman series for a process under study. On the RHS (right-hand side0 are two versions of resumming this original series. One, labeled exact, is an exact re-arrangement of the original series. The other, labeled approx., only agrees with the LHS to some fixed order N in the expansion parameter $\alpha_s$. For some time now, discussion has occurred as to which version is to be preferred~\cite{frits-ichep88}. Recently, a related more general version of this  discussion occurs for quantum gravity.\par
Whether quantum gravity is even calculable in relativistic quantum field theory is a fair but difficult question? Answers vary. According to string theory~\cite{strgthy} the answer is no, the true fundamental theory entails a one-dimensional Planck scale superstring. If we accept  loop quantum gravity~\cite{lpqg} we also find that the answer is no, the fundamental theory entails a space-time foam with a Planck scale loop structure. The answer is also no in the Horava-Lifshitz theory~\cite{horva} because the fundamental theory requires Planck scale anisoptropic scaling for space and time. Kreimer~\cite{kreimer} suggests that quantum gravity is leg-renormalizable such that the answer is yes. Weinberg~\cite{wein1} suggests that quantum gravity may be asymptotically safe, with an S-matrix
that depends only on a finite number of observable parameters, due to
the presence of a non-trivial UV fixed point, with a finite dimensional critical surface; this is equivalent to an answer of yes. We would note that the authors  in Refs.~\cite{reutera,laut,reuterb,reuter3,litim,perc},
using Wilsonian~\cite{kgw} field-space exact renormalization 
group methods, obtain results which support Weinberg's  UV fixed-point. The results in Ref.~\cite{ambj} also give support to Weinberg's
asymptotic safety suggestion.\par
In what follows, the YFS~\cite{yfs,jad-wrd} version\footnote{YFS-type soft resummation and its extension to quantum gravity was also worked-out by Weinberg in Ref.~\cite{sw-sftgrav}.} of the exact example is extended to resum the Feynman series for the Einstein-Hilbert Lagrangian for quantum gravity. In conformity with the example in eq.(\ref{eq-elmtry}), the resultant  resummed theory, resummed quantum gravity (RQG), is very much better behaved in the UV compared to what one would estimate from that Feynman series.
As we show in Refs.~\cite{bw1,bw2,bw2a,bw2i} the RQG realization of quantum gravity leads to Weinberg's UV-fixed-point behavior for the dimensionless
gravitational and cosmological constants -- the resummed theory is actually UV finite. RQG and the latter results are reviewed in Section 2.\par
The RQG theory, taken together with the Planck scale inflationary~\cite{guth,linde} cosmology formulation in Refs.~\cite{reuter1,reuter2}\footnote{The authors in Ref.~\cite{sola1} also proposed the attendant 
choice of the scale $k\sim 1/t$ used in Refs.~\cite{reuter1,reuter2}.} from the 
asymptotic safety approach to quantum gravity in 
Refs.~\cite{reutera,laut,reuterb,reuter3,litim,perc}, allows us to predict~\cite{drkuniv} the cosmological constant $\Lambda$. The prediction's closeness to the observed value~\cite{cosm1,pdg2008} motivates us to discuss its reliability and we argue~\cite{eh-consist} that its uncertainty is at the level of a factor of ${\cal O}(10)$. Constraints on susy GUT's follow. We present the Planck scale cosmology that we use and the latter results in Section 3.\par
We note that the pioneering result of Weinberg~\cite{sw-sftgrav} on summing soft gravitons is an important point of contact for our 
approach to quantum gravity. Specifically,
in an on-shell $\alpha \rightarrow \beta$ process with transition rate $\Gamma^0_{\beta\alpha}$ without soft graviton effects, Weinberg showed that inclusion of the virtual soft
graviton effects results in the transition rate
\begin{equation}
\Gamma_{\beta\alpha} = \Gamma^0_{\beta\alpha} (\lambda/U)^B,
\end{equation}
where $\lambda$ is the infrared cutoff and $U$ is the Weinberg~\cite{sw-sftgrav} soft cutoff which defines what is meant by infrared. $B$ is given by
\begin{equation}
B=\frac{G_N}{2\pi}\sum_{n,m}\eta_n\eta_mm_nm_m\frac{1+\beta_{nm}^2}{\beta_{nm}(1-\beta_{nm}^2)^{1/2}}\ln\left(\frac{1+\beta_{nm}}{1-\beta_{nm}}\right)
\label{eq1I}
\end{equation}
where $G_N$ is Newton's constant, $\eta_n=+1 (-1)$ when particle $n$ is outgoing (incoming), respectively, and $\beta_{nm}$ is the relative velocity
$\beta_{nm}={\small\sqrt{1- \frac{m_n^2m_m^2}{\left(p_np_m\right)^2}}}$ for particles $n$ and $m$ with masses $m_n$, $m_m$ and four momenta $p_n$, $p_m$, respectively. In the 2-to-2 case where 1 and 2 are incoming, 3 and 4 are outgoing, and all masses have the same value $m$, (\ref{eq1I}) shows a growth of the damping represented by $B$
with large values of $U$ as the exponential of $-(4G_Ns/\pi)\ln2\ln(U/\lambda)$ for large values of the cms energy squared $s$ for the wide-angle case with the scattering angle at $90^o$
in the center of momentum system. In our discussion
below we recover this same type of growth of the analog of $B$ with large invariant squared masses in the context of resumming the large IR regime of quantum gravity.\par
\section{Overview of Resummed Quantum Gravity}
As the Standard Theory~\footnote{We follow D.J. Gross~\cite{djg1} and call the Standard Model the Standard Theory henceforth.} of elementary particles contains many point particles, to investigate their graviton interactions, we consider\footnote{We treat spin as an inessential complication~\cite{mlgbgr}.} the Higgs-gravition extension of the Einstein-Hilbert theory,
already studied in Refs.~\cite{rpf1,rpf2}:
\begin{equation}
\begin{split}
{\cal L}(x) &= -\frac{1}{2\kappa^2} R \sqrt{-g}
            + \frac{1}{2}\left(g^{\mu\nu}\partial_\mu\varphi\partial_\nu\varphi - m_o^2\varphi^2\right)\sqrt{-g}\\
            &= \quad \frac{1}{2}\left\{ h^{\mu\nu,\lambda}\bar h_{\mu\nu,\lambda} - 2\eta^{\mu\mu'}\eta^{\lambda\lambda'}\bar{h}_{\mu_\lambda,\lambda'}\eta^{\sigma\sigma'}\bar{h}_{\mu'\sigma,\sigma'} \right\}\\
            & + \frac{1}{2}\left\{\varphi_{,\mu}\varphi^{,\mu}-m_o^2\varphi^2 \right\} -\kappa {h}^{\mu\nu}\left[\overline{\varphi_{,\mu}\varphi_{,\nu}}+\frac{1}{2}m_o^2\varphi^2\eta_{\mu\nu}\right]\\
            &  - \kappa^2 \left[ \frac{1}{2}h_{\lambda\rho}\bar{h}^{\rho\lambda}\left( \varphi_{,\mu}\varphi^{,\mu} - m_o^2\varphi^2 \right) - 2\eta_{\rho\rho'}h^{\mu\rho}\bar{h}^{\rho'\nu}\varphi_{,\mu}\varphi_{,\nu}\right] + \cdots\;.\\
\end{split}
\label{eq1-1}
\end{equation}
$R$ is the curvature scalar,  $g$ is the determinant of the metric
of space-time $g_{\mu\nu}\equiv\eta_{\mu\nu}+2\kappa h_{\mu\nu}(x) $, and $\kappa=\sqrt{8\pi G_N}$.
We expand~\cite{rpf1,rpf2} about Minkowski space
with {\small$\eta_{\mu\nu}=\text{diag}\{1,-1,-1,-1\}$}.
$\varphi(x)$, our representative scalar field for matter,  is the physical Higgs field  and
$\varphi(x)_{,\mu}\equiv \partial_\mu\varphi(x)$.
We have introduced Feynman's notation
$\bar y_{\mu\nu}\equiv \frac{1}{2}\left(y_{\mu\nu}+y_{\nu\mu}-\eta_{\mu\nu}{y_\rho}^\rho\right)$ for any tensor $y_{\mu\nu}$\footnote{Our conventions for raising and lowering indices in the 
second line of (\ref{eq1-1}) are the same as those
in Ref.~\cite{rpf2}.}. In  (\ref{eq1-1}) and in what follows, $m_o$($m$) is
the bare (renormalized) scalar boson mass.  
We set presently the small
observed~\cite{cosm1,pdg2008} value of the cosmological constant
to zero so that our quantum graviton, $h_{\mu\nu}$, has zero rest mass in (\ref{eq1-1}).
The Feynman rules for (\ref{eq1-1}) were essentially worked out by
Feynman~\cite{rpf1,rpf2}, including the rule for the famous
Feynman-Faddeev-Popov~\cite{rpf1,ffp1a,ffp1b} ghost contribution required 
for unitarity with the fixing of the gauge
(we use the gauge in Ref.~\cite{rpf1},
$\partial^\mu \bar h_{\nu\mu}=0$).
\par
As we have shown
in Refs.~\cite{bw1,bw2,bw2a}, the large virtual IR effects
in the respective loop integrals for 
the scalar propagator in quantum general relativity  
can be resummed to the {\em exact} result
$i\Delta'_F(k)=\frac{i}{k^2-m^2-\Sigma_s(k)+i\epsilon}
=  \frac{ie^{B''_g(k)}}{k^2-m^2-\Sigma'_s+i\epsilon}$
for{\small ~~~($\Delta =k^2 - m^2$)} where {\small
$B''_g(k)=\frac{\kappa^2|k^2|}{8\pi^2}\ln\left(\frac{m^2}{m^2+|k^2|}\right)$. }      
The form for $B''_g(k)$ holds for the UV(deep Euclidean) regime\footnote{ By Wick rotation, the identification
$-|k^2|\equiv k^2$ in the deep Euclidean regime gives 
immediate analytic continuation to the result for  $B''_g(k)$
when the usual $-i\epsilon,\; \epsilon\downarrow 0,$ is appended to $m^2$.}, 
so that $\Delta'_F(k)|_{\text{resummed}}$ 
falls faster than any power of $|k^2|$. See Ref.~\cite{bw1} for the analogous result
for m=0. Here, $-i\Sigma_s(k)$ is the 1PI scalar self-energy function
so that $i\Delta'_F(k)$ is the exact scalar propagator. The residual $\Sigma'_s$ starts in ${\cal O}(\kappa^2)$.
We may drop it in calculating one-loop effects. 
When the respective analogs of $i\Delta'_F(k)|_{\text{resummed}}$\footnote{These follow from the spin independence~\cite{sw-sftgrav,bw1,wein-qft} of a particle's coupling to the graviton
in the infrared regime.} are used for the
elementary particles, all quantum 
gravity loops are UV finite~\cite{bw1,bw2,bw2a}.
\par
Specifically, extending our resummed propagator results 
to all the particles
in the ST Lagrangian and to the graviton itself,
we show in the Refs.~\cite{bw1,bw2,bw2a} that
(we use $G_N$ for $G_N(0)$) 
$G_N(k)=G_N/(1+\frac{c_{2,eff}k^2}{360\pi M_{Pl}^2}),\;\; g_*=\lim_{k^2\rightarrow \infty}k^2G_N(k^2)=\frac{360\pi}{c_{2,eff}}\cong 0.0442.$
In arriving at these results, we used the result from Refs.~\cite{bw1,bw2,bw2a} for
the denominator for the propagation of transverse-traceless
modes of the graviton ($M_{Pl}$ is the Planck mass):
$q^2+\Sigma^T(q^2)+i\epsilon\cong q^2-q^4\frac{c_{2,eff}}{360\pi M_{Pl}^2},$
where $c_{2,eff}\cong  2.56\times 10^4$ is defined in Refs.~\cite{bw1,bw2,bw2a}.
\par
For the dimensionless cosmological constant $\lambda_*$ we use the VEV of 
Einstein's equation  
$G_{\mu\nu}+\Lambda g_{\mu\nu}=-\kappa^2 T_{\mu\nu},$
in a standard notation, to isolate~\cite{drkuniv}  $\Lambda$ . 
In this way, we find the deep UV limit of $\Lambda$ then becomes, allowing $G_N(k)$
to run,
$\Lambda(k) \operatornamewithlimits{\longrightarrow}_{k^2\rightarrow \infty} k^2\lambda_*,\;
\lambda_* =-\frac{c_{2,eff}}{2880}\sum_{j}(-1)^{F_j}n_j/\rho_j^2 \cong 0.0817$
where $F_j$ is the fermion number of particle $j$, $n_j$ is the effective
number of degrees of freedom of $j$ and $\rho_j=\rho(\lambda_c(m_j))$.
$\lambda_*$ vanishes in an exactly supersymmetric theory . Here, we have used the results that
a scalar
makes the contribution to $\Lambda$ given by\footnote{We note the
use here in the integrand of $2k_0^2$ rather than the $2(\vec{k}^2+m^2)$ in Ref.~\cite{bw2i}, to be
consistent with $\omega=-1$~\cite{zeld} for the vacuum stress-energy tensor.} $\Lambda_s\cong -8\pi G_N[\frac{1}{G_N^{2}64\rho^2}]$ and that a Dirac fermion contributes $-4$ times $\Lambda_s$ to
$\Lambda$, where $\rho=\ln\frac{2}{\lambda_c}$ with $\lambda_c(j)=\frac{2m_j^2}{\pi M_{Pl}^2}$ for particle j with mass $m_j$.
\par
We note that the UV fixed-point calculated here, 
$(g_*,\lambda_*)\cong (0.0442,0.0817),$ and the estimate
$(g_*,\lambda_*)\approx (0.27,0.36)$
in Refs.~\cite{reuter1,reuter2} are similar in that in both of them
$g_*$ and $\lambda_*$ are 
positive and are less than 1 in size. 
Further discussion of the relationship between the two fixed-point predictions can be found in Refs.~\cite{bw1}.
\par
\section{\bf Review of Planck Scale Cosmology and an Estimate of $\Lambda$}
The authors in Ref.~\cite{reuter1,reuter2}, using the exact renormalization group
for the Wilsonian~\cite{kgw} coarse grained effective 
average action in field space in the Einstein-Hilbert theory, as discussed in Section 1,  
have argued that the dimensionless Newton and cosmological constants
approach UV fixed points as the attendant scale $k$ goes to infinity
in the deep Euclidean regime. This is also in agreement with what we have found in RQG. 
The contact with cosmology one may facilitate via a connection between 
the momentum scale $k$ characterizing the coarseness
of the Wilsonian graininess of the average effective action and the
cosmological time $t$.  The authors
in Refs.~\cite{reuter1,reuter2}, using this latter connection, arrive at the following extension of the standard cosmological
equations:
\begin{align}
(\frac{\dot{a}}{a})^2+\frac{K}{a^2}&=\frac{1}{3}\Lambda+\frac{8\pi}{3}G_N\rho,\nonumber\\
\dot{\rho}+3(1+\omega)\frac{\dot{a}}{a}\rho&=0,\;
\dot{\Lambda}+8\pi\rho\dot{G_N}=0,\nonumber\\
G_N(t)&=G_N(k(t)),\;
\Lambda(t)=\Lambda(k(t)).
\label{coseqn1}
\end{align}
Here, $\rho$ is the density and $a(t)$ is the scale factor
with the Robertson-Walker metric given as
\begin{equation}
ds^2=dt^2-a(t)^2\left(\frac{dr^2}{1-Kr^2}+r^2(d\theta^2+\sin^2\theta d\phi^2)\right)
\label{metric1}
\end{equation}
where $K=0,1,-1$ correspond respectively to flat, spherical and
pseudo-spherical 3-spaces for constant time t.  
The attendant equation of state is 
$ 
p(t)=\omega \rho(t),
$
where $p$ is the pressure.
The aforementioned relationship between $k$ and the cosmological time $t$ is 
$
k(t)=\frac{\xi}{t}
$
with the constant $\xi>0$ determined
from constraints on
physical observables.\par 
Using the UV fixed points for $k^2G_N(k)\equiv g_*$ and
$\Lambda(k)/k^2\equiv \lambda_*$ obtained independently, the authors in Refs.~\cite{reuter1,reuter2} solve the cosmological system in Eqs.(\ref{coseqn1}). They find, for $K=0$,
a solution in the Planck regime where $0\le t\le t_{\text{class}}$, with
$t_{\text{class}}$ a ``few'' times the Planck time $t_{Pl}$, which joins
smoothly onto a solution in the classical regime, $t>t_{\text{class}}$,
which coincides with standard Friedmann-Robertson-Walker phenomenology
but with the horizon, flatness, scale free Harrison-Zeldovich spectrum,
and entropy problems all solved purely by Planck scale quantum physics.
We now recapitulate how to use the Planck scale cosmology of Refs.~\cite{reuter1,reuter2} and the UV limits $ \{g_*,\; \lambda_*\}$ in RQG~\cite{bw1,bw2,bw2a} in Refs.~\cite{bw2i} to predict~\cite{drkuniv} the current value of $\Lambda$.
\par
Specifically, the transition time between the Planck regime and the classical Friedmann-Robertson-Walker(FRW) regime is determined as $t_{tr}\sim 25 t_{Pl}$ in the Planck scale cosmology description of inflation in Ref.~\cite{reuter2}. 
In Ref.~\cite{drkuniv} we show that, starting with the quantity $\rho_\Lambda(t_{tr}) \equiv\frac{\Lambda(t_{tr})}{8\pi G_N(t_{tr})}$, we get, 
following the arguments in Refs.~\cite{branch-zap} ($t_{eq}$ is the time of matter-radiation equality),  
\begin{equation}
\begin{split}
\rho_\Lambda(t_0)&\cong \frac{-M_{Pl}^4(1+c_{2,eff}k_{tr}^2/(360\pi M_{Pl}^2))^2}{64}\sum_j\frac{(-1)^Fn_j}{\rho_j^2}
          \times \frac{t_{tr}^2}{t_{eq}^2} \times (\frac{t_{eq}^{2/3}}{t_0^{2/3}})^3\cr
   & \cong \frac{-M_{Pl}^2(1.0362)^2(-9.194\times 10^{-3})}{64}\frac{(25)^2}{t_0^2}
   \cong (2.4\times 10^{-3}eV)^4.\cr
\end{split}
\label{eq-rho-expt}
\end{equation}
$t_0\cong 13.7\times 10^9$ yrs. is the age of the universe. 
The estimate in (\ref{eq-rho-expt}) is close to the experimental result~\cite{pdg2008}\footnote{The analysis in Ref.~\cite{sola2} also gives 
a value for $\rho_\Lambda(t_0)$ that is qualitatively similar to this experimental result.} 
$\rho_\Lambda(t_0)|_{\text{expt}}\cong ((2.37\pm 0.05)\times 10^{-3}eV)^4$. 
\par
In Ref.~\cite{drkuniv}, detailed discussions are given of the three issues of the effect of various spontaneous symmetry breaking energies on $\Lambda$, the effect of our approach to $\Lambda$ on big bang nucleosynthesis(BBN)~\cite{bbn}, and the effect of the time dependence of $\Lambda$ and $G_N$ on the covariance~\cite{bianref1,bianref2,bianref3} of the theory. We refer the reader to the respective discussions in Ref.~\cite{drkuniv}. In Ref.~\cite{eh-consist}, we have argued, regarding the issue of the error on our estimate, that the structure of the solutions of Einstein's equation, taken together with the Heisenberg uncertainty principle,  
implies the constraint $
k\ge \frac{\sqrt{5}}{2w_0}=\frac{\sqrt{5}}{2}\frac{1}{\sqrt{3/\Lambda(k)}}
$
where $\Lambda(k)$  follows from (\ref{eq-rho-expt}) (see  Eq.(52) in Ref.~\cite{drkuniv}). This constraint's equality gives the estimate~\cite{drkuniv}
of the transition time, $t_{\text tr}=\alpha/M_{Pl}=1/k_{\text{tr}}$, from the Planck scale inflationary regime~\cite{reuter1,reuter2} to the Friedmann-Robertson-Walker regime via the implied value of $\alpha$. On solving this equality for $\alpha$ we get $
\alpha\cong 25.3,$
in agreement with the value $\alpha\cong 25$ implied by the numerical studies in Ref.~\cite{reuter1,reuter2}. This agreement suggests an error on $t_{\text tr}$ at the level of a factor ${\cal O}(3)$ or less and an uncertainty on $\Lambda$ reduced from a factor of ${\cal O}(100)$~\cite{drkuniv} to a factor of ${\cal O}(10)$.\par
One may ask what would happen to our estimate if there were a susy GUT theory at high scales? Even though the LHC has yet to see~\cite{susylmt} any trace of susy, it may still appear. In Ref.~\cite{drkuniv}, for definiteness and purposes of illustration, 
we use the susy SO(10) GUT model in Ref.~\cite{ravi-1}
to illustrate how such a theory might affect our estimate of $\Lambda$. We show that either one needs a very high mass for the gravitino or one needs twice the usual particle content with the susy partners of the new quarks and leptons at masses much lower than their partners' masses -- see Ref.~\cite{drkuniv}.\par

\setlength{\bibsep}{1.7pt}

\end{document}